\documentstyle[12pt]{article}
\begin{document}
\begin{center}

{\Large \bf BERRY PHASE CONNECTION AND ANOMALY\\
            IN TWO DIMENSIONS}\footnote{To be
presented at the Workshop of the Research Program
"Quantization, Generalized BRS Cohomology and Anomalies",
September 1998, Erwin Schr\"{o}dinger Institute,
Vienna, Austria}\\

\vspace{2 cm}

{\large \bf Fuad M. Saradzhev}\footnote{Address after
October 1st : Centre for Theoretical Physics, University
of Sussex, Brighton, UK}

\vspace{1 cm}

{\small \it Institute of Physics, Academy of Sciences of Azerbaijan,\\
\it Huseyn Javid pr. 33, 370143 Baku, AZERBAIJAN}\footnote{Permanent
address}\\

{\small \it and}\\

{\small \it The Abdus Salam International Centre for
Theoretical Physics,\\
Strada Costiera 11, P.O. Box 586, 34100 Trieste, ITALY}\\

\end{center}

\vspace{3 cm}

\begin{center}

{\bf ABSTRACT}

\end{center}

\vspace{1 cm}

\rm

For the generalized chiral Schwinger model defined on the circle,
a direct calculation of the zero curvature part of the vacuum
Berry phase connection is given. Although this part does not
contribute to the curvature, it is shown to attach several
features to the total connection and to produce a physical
background of linearly rising electric fields.

\newpage
\rm
  
1. Berry phase \cite{berry84} plays an important role
in gauge models with anomaly \cite{bert96}. 
A common topological nature
of this phase and gauge anomalies was shown in \cite{nels85}.
It was proved that a $U(1)$ connection related to the
vacuum Berry phase contributes to the Hamiltonian and
that just this contribution makes the theory gauge
invariant \cite{niemi85,niemi86}. In \cite{sara97} an interrelation
between the nonvanishing vacuum Berry phase and anomaly
was demonstrated explicitly for the generalized chiral
Schwinger model ( or chiral $QED_2$ ).

Usually the Berry phase connection is not calculated
directly. It is not well defined globally on a manifold
of all static gauge field configurations and is not
invariant under gauge field dependent redefinitions
of the phases of the states acquiring the Berry phase . 
That is why we first compute the
corresponding $U(1)$ curvature tensor and then deduce
from it an expression for the connection.

However, in this way the Berry phase connection is not
determined uniquely. We can not fix that part of the connection
which has vanishing curvature. Namely, if ${\cal A}$ is
a connection deduced from the curvature ${\cal F}$, then
${\cal A} + {\cal A}^{(0)}$ where ${\cal A}^{(0)}$
is an arbitrary connection with zero curvature also
corresponds to ${\cal F}$. Since the vacuum Berry phase
connection represents a background in which the physical
degrees of freedom of anomalous models are moving, a
direct calculation of its zero curvature part is very
important for understanding the dynamics of such models.

In the present note, we aim to calculate the zero curvature
part of the Berry phase connection for the vacuum
Fock state in the generalized
chiral $QED_2$ defined on the circle $S^1$. We work in the temporal 
gauge $A_0=0$ and use the system of
units where $c=1$. In the generalized chiral $QED_2$ a $U(1)$ 
gauge field is coupled with different charges to both
chiral components of a fermionic field. Only matter fields
are quantized, while the gauge field is handled as a
classical field. We show that the zero curvature
Berry phase connection is nothing else than the linearly
rising electric field found previously in \cite{sarad92,fm97}.

\vspace{1 cm}

2. In the temporal gauge $A_0=0$, the fermionic Hamiltonian density 
of the generalized chiral $QED_2$ is \cite{sara97,fm97}
\[
{\cal H}_{\rm F} = {\cal H}_{+} + {\cal H}_{-} ,
\]
\begin{equation}
{\cal H}_{\pm}  \equiv  \psi_{\pm}^{\dagger} d_{\pm} \psi_{\pm} =
\mp \psi_{\pm}^{\dagger}(i{\hbar}{\partial}_{1}+e_{\pm}A_1)\psi_{\pm},
\label{eq: odin}
\end{equation}
where ${\psi}_{+}$ and ${\psi}_{-}$ are correspondingly positive
and negative chirality matter fields.
We suppose that space is a circle of length $L$,
$-\frac{L}{2} \leq x < \frac{L}{2}$ , and that the fields
are periodic on the circle.

The eigenfunctions and the eigenvalues of the first quantized
fermionic Hamiltonians are
\[
d_{\pm} \langle x|n;{\pm} \rangle = \pm \varepsilon_{n,{\pm }}
\langle x|n;{\pm } \rangle ,
\]
where
\[ 
\langle x|n;{\pm } \rangle = \frac{1}{\sqrt {\rm L}}
\exp\{\frac{i}{\hbar} e_{\pm} \int_{-{\rm L}/2}^{x} dz{A_1}(z) +
\frac{i}{\hbar} \varepsilon_{n,{\pm}} \cdot x\},  
\]
\[
\varepsilon_{n,{\pm }} = \frac{2\pi}{\rm L} 
(n{\hbar} - \frac{e_{\pm}b{\rm L}}{2\pi}),
\]
and
\[
b \equiv \frac{1}{L} \int_{-L/2}^{L/2} dx A_1(x)
\]
is the gauge field zero mode.

Now we introduce the second quantized positive and
negative chirality Dirac fields. 
At time $t=0$, in terms of the 
eigenfunctions of the first quantized fermionic Hamiltonians the 
second 
quantized ($\zeta$--function regulated) fields have the expansion :
\[
\psi_{+}^s (x) = \sum_{n \in \cal Z} a_n \langle x|n;{+} \rangle
|\lambda \varepsilon_{n,+}|^{-s/2},
\]
\begin{equation}
\psi_{-}^s (x) = \sum_{n \in \cal Z} b_n \langle x|n;{-} \rangle
|\lambda \varepsilon_{n,-}|^{-s/2}.
\label{eq: dva}
\end{equation}
Here $\lambda$ is an arbitrary constant with dimension of length
which is necessary to make $\lambda \varepsilon_{n,\pm}$ 
dimensionless,
while $a_n, a_n^{\dagger}$ and $b_n, b_n^{\dagger}$ are 
correspondingly 
positive and negative chirality fermionic annihilation and creation
operators which fulfil the commutation relations
\[  
[a_n , a_m^{\dagger}]_{+} = [b_n , b_m^{\dagger}]_{+} =\delta_{m,n} .
\]

The vacuum state 
\[
|{\rm vac};A \rangle = |{\rm vac};A;+ \rangle \otimes
|{\rm vac};A;- \rangle 
\]
is defined such that all negative energy
levels are filled and the others are empty:
\begin{eqnarray}
a_n|{\rm vac};A;+\rangle =0 & {\rm for} &
n>[\frac{e_{+}b{\rm L}}{2{\pi}{\hbar}}],
\nonumber \\
a_n^{\dagger} |{\rm vac};A;+ \rangle =0 & {\rm for} & n \leq
[\frac{e_{+}b{\rm L}}{2{\pi}{\hbar}}],
\label{eq: tri}
\end{eqnarray}
and
\begin{eqnarray}
b_n|{\rm vac};A;-\rangle =0 & {\rm for} & n \leq
[\frac{e_{-}b{\rm L}}{2{\pi}{\hbar}}], \nonumber \\
b_n^{\dagger} |{\rm vac};A;- \rangle =0 & {\rm for} & n >
[\frac{e_{-}b{\rm L}}{2{\pi}{\hbar}}],
\label{eq: cet}
\end{eqnarray}
where $[\frac{e_{\pm}bL}{2{\pi}\hbar}]$ are integer parts
of $\frac{e_{\pm}bL}{2{\pi}\hbar}$.

Next we define the fermionic parts 
of the second-quantized Hamiltonian as
\[
\hat{\rm H}_{\pm}^s = \int_{-{\rm L}/2}^{{\rm L}/2} dx
\hat{\cal H}_{\pm}^s(x)= \frac{1}{2} \int_{-{\rm L}/2}^{{\rm L}/2}dx
 (\psi_{\pm}^{\dagger s} d_{\pm} \psi_{\pm}^s
- \psi_{\pm}^s d_{\pm}^{\star} \psi_{\pm}^{\dagger s}).
\]
Substituting (\ref{eq: dva}) into these expressions, we obtain
\begin{eqnarray*}
\hat{\rm H}_{+} & = & \lim_{s \to 0}
\sum_{k \in \cal Z} \varepsilon_{k,+} a_k^{\dagger}
a_k
|\lambda \varepsilon_{k,+}|^{-s}  , \\
\hat{\rm H}_{-} & = & \lim_{s \to 0}
\sum_{k \in \cal Z} \varepsilon_{k,-} b_{k} b_{k}^
{\dagger}
|\lambda \varepsilon_{k,-}|^{-s} .
\end{eqnarray*}
The operators $\hat{\rm H}_{\pm}$ are
well defined when acting on finitely excited states which have only a
finite number of excitations relative to the Fock vacuum.

\vspace{1 cm}

3. In the adiabatic approach \cite{schi68,zwan90}, 
the dynamical variables are divided
into two sets, one which we call fast variables and the other
which we call slow variables. In our case, we treat the fermions
as fast variables and the gauge fields as slow variables.

Let ${\cal A}^1$ be a manifold of all static gauge field
configurations ${A_1}(x)$. On ${\cal A}^1$  a time-dependent
gauge field corresponds to a path and a periodic gauge
field to a closed loop.

The second-quantized fermionic Hamiltonian
$:\hat{\rm H}_{\rm F}:=:\hat{\rm H}_{+}: + :\hat{\rm H}_{-}:$
normal ordered with respect to the vacuum state depends 
on $t$  through the background
gauge field $A_1$ and so changes very slowly with time.
Let us assume that the background gauge field
${A_1}(x,t)$ is periodic $(0 \leq t <T)$~. After a
time $T$ the periodic field returns to its original
value: ${A_1}(x,0) = {A_1}(x,T)$, so that $:\hat{\rm H}_{\pm}:(0)=
:\hat{\rm H}_{\pm}:(T)$ .

At each instant $t$ we define eigenstates for $:\hat{\rm H}_{\pm}:
(t)$ by
\[
:\hat{\rm H}_{\pm}:(t) |{\rm F}, A(t);\pm \rangle =
{\varepsilon}_{{\rm F};\pm}(t) |{\rm F}, A(t);\pm \rangle.
\]
The Fock states $|{\rm F}, A(t) \rangle = |{\rm F}, A(t); + \rangle
\otimes |{\rm F}, A(t); - \rangle$ depend on $t$ only through
their implicit dependence on $A_1$. They are assumed to be
orthonormalized and nondegenerate.

According to the adiabatic approximation, upon parallel
transport around a closed loop on ${\cal A}^1$ the
Fock states $|{\rm F};A(t);\pm \rangle$ can acquire
only a phase. This phase consists of two parts , the
usual dynamical phase and an extra phase discovered by
Berry. Whereas the dynamical phase provides information
about the duration of the evolution, the Berry's phase
\[
{\gamma}_{{\rm F},\pm}^{\rm Berry} =  
\int_{0}^{T} dt \int_{-{\rm L}/2}^
{{\rm L}/2} dx \dot{A_1}(x,t) {\cal A}_{{\rm F},\pm}(x,t),
\]
which is integrated exponential
of the $U(1)$ connections
\begin{equation}
{\cal A}_{{\rm F};\pm}(x,t) = \langle {\rm F},A(t); \pm
|i \frac{\delta}
{\delta A_1(x,t)}|{\rm F},A(t),\pm \rangle,
\label{eq: pet}
\end{equation}
reflects the nontrivial holonomy of the Fock states
on ${\cal A}^1$.

The connections ${\cal A}_{{\rm F},\pm}$ can be defined
only locally on ${\cal A}^1$, in regions where
$[\frac{e_{\pm}bL}{2{\pi}{\hbar}}]$ are fixed. If
$[\frac{e_{\pm}bL}{2{\pi}{\hbar}}]$ change, then there is
a nontrivial spectral flow, i.e. some energy levels of
the first quantized fermionic Hamiltonians cross zero
and change sign. This means that the definition
of the Fock vacuum of the second quantized
fermionic Hamiltonian changes. Since the creation and
annihilation operators $a^{\dagger} , a$ ( and $b^{\dagger} ,
b )$ are continuous functionals of $A_1(x)$, the definition
of all excited Fock states is also discontinuous. The
connections ${\cal A}_{{\rm F},\pm}$ are not therefore
well defined globally.

Moreover, ${\cal A}_{{\rm F},\pm}$ are not invariant under
$A$-dependent redefinitions of the phases of the Fock
states. For these reasons,
we usually compute first the $U(1)$ curvature tensors
\[
{\cal F}_{\rm F}^{\pm}(x,y,t) \equiv
\frac{\delta}{\delta A_1(x,t)}
{\cal A}_{{\rm F},\pm}(y,t) - \frac{\delta}{\delta A_1(y,t)}
{\cal A}_{{\rm F},\pm}(x,t)
\]
and then deduce ${\cal A}_{{\rm F},\pm}$.

In particular, for the vacuum states the curvature tensors
are \cite{sara97}
\[
{\cal F}_{{\rm vac}}^{\pm}(x,y,t) = \pm
\frac{e_{\pm}^2}{2{\pi}{\hbar}^2}
( \frac{1}{2} \epsilon(x-y)
- \frac{1}{\rm L} (x-y) ),
\]
so the corresponding connections are deduced as
\begin{equation}
{\cal A}_{{\rm vac},\pm}(x,t) = {\cal A}_{0,\pm}(x,t)
-\frac{1}{2} \int_{-{\rm L}/2}
^{{\rm L}/2} dy {\cal F}_{{\rm vac}}^{\pm}(x,y,t) A_1(y,t),
\label{eq: shest}
\end{equation}
where ${\cal A}_{0,\pm}(x,t)$ are arbitrary connections
which have zero curvature and can not be therefore fixed
by this procedure . In \cite{sara97} we put
${\cal A}_{0,\pm}(x,t)=0$.

Let us show that the zero curvature part of the
connections (\ref{eq: shest}) can be computed directly
by using a Fock vacuum whose definition is globally single-valued .
We introduce the new Fock vacuum $\overline{|{\rm vac};A
\rangle}=
\overline{|{\rm vac};A;+ \rangle} \bigotimes
\overline{|{\rm vac};A;- \rangle}$
defined as
\begin{eqnarray*}
a_n \overline{|{\rm vac};A;+ \rangle}=0
& {\rm for} & n>0, \\
a_n^{\dagger} \overline{|{\rm vac};A;+ \rangle}=0
& {\rm for} & n \leq 0,
\end{eqnarray*}
and
\begin{eqnarray*}
b_n \overline{|{\rm vac};A;- \rangle}=0
& {\rm for} & n \leq 0, \\
b_n^{\dagger} \overline{|{\rm vac};A;- \rangle}=0
& {\rm for} & n>0.
\end{eqnarray*}
The new Fock vacuum is defined such that
for all values of $[\frac{e_{\pm}b{\rm L}}{2\pi\hbar}]$ only the
levels with energy lower than (or equal to) the energy of the
level $n=0$ are filled and the others are empty, i.e. the new
definition does not depend on $[\frac{e_{\pm}b{\rm L}}{2\pi\hbar}]$
and remains unchanged as the gauge configuration changes.
The definition of all excited states constructed over the new Fock vacuum is
also globally single-valued.

In the region where $[\frac{e_{\pm}bL}{2{\pi}{\hbar}}]=0$
the old and the new Fock vacuums coincide,
\[
|{\rm vac};A;\pm \rangle_{(0)} =
\overline{|{\rm vac};A;\pm \rangle},
\]
where the subscript $(0)$ indicates that
$[\frac{e_{\pm}bL}{2{\pi}{\hbar}}]$ vanish.

In regions with nonzero $[\frac{e_{\pm}bL}{2{\pi}{\hbar}}]$
the positive chirality vacuums are connected as
\[
|{\rm vac};A;+ \rangle_{(k_{+})} =
a_{k_{+}}^{\dagger} \cdot . . . \cdot a_2^{\dagger}
a_1^{\dagger} \overline{|{\rm vac};A;+ \rangle}
\]
for $[\frac{e_{+}bL}{2{\pi}{\hbar}}] \equiv k_{+} >0$,
and
\[
|{\rm vac};A;+ \rangle_{(k_{+})} = a_{k_{+}+1} \cdot
. . . \cdot a_{-1} a_{0} \overline{|{\rm vac};A;+ \rangle}
\]
for $k_{+}<0$.

For $k_{+}=\pm 1$, the old and the new Fock vacuums of
positive chirality are compared in Figures 1 and 2.

In the negative chirality sector we have analogously
\[
|{\rm vac};A;- \rangle_{(k_{-})} = b_{k_{-}} \cdot
. . . \cdot b_{2} b_1 \overline{|{\rm vac};A;- \rangle}
\]
for $[\frac{e_{-}bL}{2{\pi}{\hbar}}] \equiv k_{-} >0$,
and
\[
|{\rm vac}; A; - \rangle_{(k_{-})} =b_{k_{-}+1}^{\dagger}
\cdot . . . \cdot b_{-1}^{\dagger} b_{0}^{\dagger}
\overline{|{\rm vac};A;- \rangle}
\]
for $k_{-}<0$.

Next we define the vacuum Berry phase connections in the
regions of different values of $k_{+}$:
\[
{\cal A}_{{\rm vac},+}^{(k_{+},q_{+})}(x,t) \equiv
{ _{(k_{+})}\langle} {\rm vac};A;+|i \frac{\delta}
{\delta A_1(x,t)}|{\rm vac};A;+ \rangle_{(q_{+})}.
\]
For simplicity, we start with the positive chirality
sector and positive values of $k_{+}$,$q_{+}$.

Making transition to the new Fock vacuum, we get
\begin{equation}
{\cal A}_{{\rm vac},+}^{(k_{+},q_{+})}(x,t) =
\overline{\langle {\rm vac};A;+|}
a_1 a_2 \cdot ... \cdot a_{k_{+}} \cdot i \frac{\delta}
{\delta A_1(x,t)} \cdot a_{q_{+}}^{\dagger} \cdot ...
\cdot a_2^{\dagger} a_1^{\dagger}
\overline{|{\rm vac};A;+ \rangle}.
\label{eq: sem}
\end{equation}
For $k_{+} \neq q_{+}$, the number of creation operators
in (\ref{eq: sem}) is not equal to the number of
annihilation ones. Those creation (or annihilation)
operators which have not their annihilation (or creation)
counterparts annihilate the vacuum state
$\overline{\langle {\rm vac};A;+|}$ ( or
$\overline{|{\rm vac};A;+ \rangle}$ ), so that the
connections ${\cal A}_{{\rm vac},+}^{(k_{+},q_{+})}$
vanish.

For $k_{+}=q_{+}$, all the creation and annihilation
operators in (\ref{eq: sem}) can be paired. In the
$\zeta$-function regularization scheme, the action
of the functional derivative ${\delta}/{\delta A_1}$
on the operators $a_n$, $a_n^{\dagger}$ is given by
\cite{rept86}
\begin{eqnarray*}
\frac{\delta}{\delta A_1} a_n & = & - \lim_{s \to 0}
\sum_{m \in \cal Z} \langle n;+|\frac{\delta}
{\delta A_1}|m;+ \rangle a_m |{\lambda}
{\varepsilon}_{m,+}|^{-s/2}, \\
\frac{\delta}{\delta A_1} a_n^{\dagger} & = & \lim_{s \to 0}
\sum_{m \in \cal Z} \langle m;+|\frac{\delta}
{\delta A_1}|n;+ \rangle a_m^{\dagger} |{\lambda}
{\varepsilon}_{m,+}|^{-s/2}.\\
\end{eqnarray*}

Eq.(\ref{eq: sem}) is then rewritten as
\[
{\cal A}_{{\rm vac},+}^{(k_{+},k_{+})} =
\overline{\cal A}_{{\rm vac},+} + \lim_{s \to 0}
\sum _{n=1}^{k_{+}} \langle n;+|i \frac{\delta}
{\delta A_1}|n;+ \rangle |{\lambda} {\varepsilon}_{n,+}|^{-s/2},
\]
where
\[
\overline{\cal A}_{{\rm vac};+}(x,t) \equiv
\overline{\langle {\rm vac};A;+|} i \frac{\delta}
{\delta A_1(x,t)} \overline{|{\rm vac};A;+ \rangle}
\]
is the Berry phase connection for the new vacuum
of positive chirality.

By a direct calculation of the expectation values of
$i \frac{\delta}{\delta A_1}$ for the first
quantized kets and bras we obtain
\begin{equation}
{\cal A}_{{\rm vac};+}^{(k_{+},k_{+})} =
\overline{\cal A}_{{\rm vac},+} + \frac{1}{\hbar}
\frac{e_{+}}{\rm L} (x-\frac{\rm L}{2}) k_{+}.
\label{eq: vosem}
\end{equation}

For negative values of $k_{+}$,$q_{+}$, we get the same
result, namely, if $k_{+} \neq q_{+}$ including the cases
$k_{+}>0$, $q_{+}<0$ and $k_{+}<0$, $q_{+}>0$, then
${\cal A}_{{\rm vac},+}^{(k_{+},q_{+})}$ vanish,
while for $k_{+}=q_{+}$ the connections are given by
Eq.(8).

In the negative chirality sector, the connections
\[
{\cal A}_{{\rm vac},-}^{(k_{-},q_{-})}(x,t) \equiv
{ _{(k_{-})}\langle} {\rm vac};A;-|i\frac{\delta}{\delta A_1(x,t)}|
{\rm vac};A;- \rangle_{(q_{-})}
\]
are computed analogously, being nonzero as before only
for $k_{-}=q_{-}$ :
\begin{equation}
{\cal A}_{{\rm vac},-}^{(k_{-},k_{-})} =
\overline{\cal A}_{{\rm vac},-} - \frac{1}{\hbar}
\frac{e_{-}}{\rm L} (x - \frac{\rm L}{2}) k_{-},
\label{eq: devet}
\end{equation}
where
\[
\overline{\cal A}_{{\rm vac},-} \equiv
\overline{\langle {\rm vac};A;-|} i\frac{\delta}
{\delta A_1} \overline{|{\rm vac};A;- \rangle}.
\]

Using the definition of $k_{\pm}$, we immediately
reproduce from Eqs.(8) and (9) the vacuum Berry
connections defined on whole manifold ${\cal A}^1$:
\begin{equation}
{\cal A}_{{\rm vac};\pm} =
\overline{\cal A}_{{\rm vac};\pm} \pm \frac{1}{\hbar}
\frac{e_{\pm}}{\rm L} (x - \frac{\rm L}{2})
[\frac{e_{\pm}b{\rm L}}{2{\pi}{\hbar}}].
\label{eq: deset}
\end{equation}

Only the first term in (\ref{eq: deset}) contributes to
the vacuum curvature tensor, ${\cal F}_{\rm vac}^{\pm}=
\overline{\cal F}_{\rm vac}^{\pm}$, and can be therefore
deduced from the curvature, so we write
\begin{equation}
\overline{\cal A}_{{\rm vac},\pm}(x,t) = - \frac{1}{2}
\int_{-{\rm L}/2}^{{\rm L}/2} dy {\cal F}_{\rm vac}^{\pm}
(x,y,t) A_1(y,t).
\label{eq: odinodin}
\end{equation}

The second term has vanishing curvature, and we identify
it with ${\cal A}_{0,\pm}(x,t)$ :
\begin{equation}
{\cal A}_{0,\pm}(x,t) = \pm \frac{1}{\hbar} \frac{e_{\pm}}{\rm L}
(x - \frac{\rm L}{2}) [\frac{e_{\pm}b{\rm L}}{2{\pi}{\hbar}}].
\label{eq: odindva}
\end{equation}

We see that just the zero curvature part of the vacuum
Berry phase connections is discontinuous on ${\cal A}^1$.
In the transition between regions of different
$[\frac{e_{\pm}b{\rm L}}{2{\pi}{\hbar}}]$ the zero
curvature part changes by a multiple of
$ \pm \frac{1}{\hbar} \frac{e_{\pm}}{\rm L} (x - \frac{\rm L}{2})$.

In the models with anomaly the vacuum Berry connection is added
to the electric field operator in order to keep gauge invariant
the full quantum theory with both matter and gauge fields
quantized \cite{niemi85}. This connection represents a
background for the quantum physical degrees of freedom. In
our case, the background corresponding to the zero 
curvature connections (\ref{eq: odindva}) consists
of linearly rising electric fields
\[
{\cal E}_{\pm}(x) = \mp \frac{e_{\pm}}{\rm L} x
[\frac{e_{\pm}b{\rm L}}{2{\pi}{\hbar}}],
\] 
so that
\[
{\cal A}_{0,\pm}(x,t) = \frac{1}{\hbar} \left( {\cal E}_{\pm}(\frac{\rm L}{2})
- {\cal E}_{\pm}(x)  \right) .
\]

\vspace{1 cm}

4. In conclusion, we have calculated the zero curvature part
of the vacuum Berry phase connections directly by using
the globally single-valued definition for the Fock vacuum.

The zero curvature vacuum Berry connections have several features.
They depend only on the zero mode of the gauge field, change
discontinuously on ${\cal A}^1$ and define a background of
linearly rising electric fields.

That part of the vacuum Berry phase connections which contributes
to the curvature is continuous on ${\cal A}^1$ and depends on
the gauge field non-zero modes. For any gauge field defined
on the circle, its zero mode represents the only global
physical degree of freedom, while the non-zero ones are
gauge variant and can be removed by the gauge transformations.
The background defined by the zero curvature part of the
Berry phase connections is therefore a physical one and
survives in the physical sector of the full quantum theory.
This indicates that the zero curvature vacuum Berry connections
play an essential role in the construction of the physical
quantum picture of the anomalous models.

For the chiral Schwinger model, the existence of the
background of the linearly rising electric field
${\cal E}(x) = {\cal E}_{+}(x) + {\cal E}_{-}(x)$
was proved previously in \cite{sarad92,fm97}.
In the present paper, we have obtained the same
background in the framework of the adiabatic approximation.

It would be of interest to calculate directly the
connections $\overline{\cal A}_{{\rm vac},\pm}$ as well.
Then we could check whether the separation of the zero
and non-zero curvature parts in (\ref{eq: deset}) is
complete or not.

Both the zero and non-zero curvature parts of the vacuum
Berry phase connections are associated with anomaly.
The motion of the physical degrees of freedom in the
linearly rising background electric field represents a new
type of interactions which are absent in the nonanomalous models
\cite{fm97}. For the standard Schwinger model with $e_{+}=e_{-}$,
the zero curvature parts of the Berry phase connections for
the vacuums of positive and negative chiralities are 
opposite in sign and so cancel each other.
The total non-zero curvature part of the vacuum Berry connections
is related to $1$-cocycle of the gauge group projective
representation responsible for anomaly and also vanishes
for $e_{+}=e_{-}$ \cite{sara97}.

\newpage

\vspace{1 cm}

{\bf Acknowledgement}

\vspace{5 mm}

\begin{flushleft}

The author would like to thank S. Azakov for useful discussions
and comments.

\end{flushleft}

\newpage

\vspace{1 cm}

\begin{center}

{\large \bf Figure Captions}

\end{center}

\vspace{2 cm}

{\bf FIG. 1.} Schematic representation of vacuum states
for $[\frac{e_{+}b{\rm L}}{2{\pi}{\hbar}}]=1$ :
(a): $|{\rm vac};A;+\rangle$ ,\\
(b): $\overline{|{\rm vac};A;+\rangle}$.
Only the positive chirality sector is shown.

\vspace{2 cm}

{\bf FIG. 2.} Schematic representation of vacuum states
for $[\frac{e_{+}b{\rm L}}{2{\pi}{\hbar}}]=-1$ :
(a): $|{\rm vac};A;+\rangle$ ,\\
(b): $\overline{|{\rm vac};A;+\rangle}$.
Only the positive chirality sector is shown.

\newpage

\vspace{3 cm}

\setlength{\unitlength}{1cm}

\begin{picture}(18,12)(-9,0)
{\thicklines
\multiput (-7,4)(0,.5){6}{\line(1,0){1}}}
\multiput (-6.5,7.5)(0,.5){6}{\circle*{.1}} 
\put (-6.5,11){\vector(0,1){1}}
\put (-6.5,3){\vector(0,-1){1}}
\thinlines
\multiput (-9,7)(1,0){5}{\line(1,0){.5}}
\put (-5,4.9){n=-2}
\put (-5,5.4){n=-1}
\put (-5,5.9){n=0}
\put (-5,6.4){n=1}
\put (-5,7.4){n=2}
\put (-6.7,1){\bf (a)}
\multiput (-2,7)(1,0){5}{\line(1,0){.5}}
\multiput (.5,7.5)(0,.5){6}{\circle*{.1}}
\put (.5,6.5){\circle*{.1}}
{\thicklines
\multiput (0,4)(0,.5){5}{\line(1,0){1}}}
\put (2,4.9){n=-2}
\put (2,5.4){n=-1}
\put (2,5.9){n=0}
\put (2,6.4){n=1}
\put (2,7.4){n=2}
\put (.2,1){\bf (b)}
\put (.5,3){\vector(0,-1){1}}
\put (.5,11){\vector(0,1){1}}
\put (4.5,11){\vector(0,1){1}}
\put (5,11){\rm Energy}
\put (4.5,9){\circle*{.1}}
\put (6,9){\rm empty level}
{\thicklines
\put (4,8.5){\line(1,0){1}}}
\put (6,8.5){\rm filled level}
\multiput (3.8,8)(.6,0){3}{\line(1,0){.3}}
\put (6,8){\rm fermi surface}
\end{picture}

\vspace{1 cm}

\begin{center}

{\bf FIG. 1.}

\end{center}

\newpage

\vspace{3 cm}

\begin{picture}(18,12)(-9,0)
{\thicklines
\multiput (-7,4)(0,.5){6}{\line(1,0){1}}}
\multiput (-6.5,7.5)(0,.5){6}{\circle*{.1}}
\put (-6.5,11){\vector(0,1){1}}
\put (-6.5,3){\vector(0,-1){1}}
\thinlines
\multiput (-9,7)(1,0){5}{\line(1,0){.5}}
\put (-5,5.9){n=-2}
\put (-5,6.4){n=-1}
\put (-5,7.4){n=0}
\put (-5,7.9){n=1}
\put (-5,8.4){n=2}
\put (-6.7,1){\bf (a)}
\multiput (-2,7)(1,0){5}{\line(1,0){.5}}
\multiput (.5,8)(0,.5){5}{\circle*{.1}}
{\thicklines
\put (0,7.5){\line(1,0){1}}
\multiput (0,4)(0,.5){6}{\line(1,0){1}}}
\put (2,5.9){n=-2}
\put (2,6.4){n=-1}
\put (2,7.4){n=0}
\put (2,7.9){n=1}
\put (2,8.4){n=2}
\put (.5,3){\vector(0,-1){1}}
\put (.5,11){\vector(0,1){1}}
\put (.2,1){\bf (b)}
\put (4.5,11){\vector(0,1){1}}
\put (5,11){\rm Energy}

\end{picture}

\vspace{1 cm}

\begin{center}

{\bf FIG. 2.}

\end{center}

\end{document}